\documentclass[runningheads]{llncs}
\usepackage{graphicx}
\usepackage{subfigure}
\usepackage{hyperref}
\usepackage[dvipsnames]{xcolor}

\begin{document}

\title{Walk4Me: Telehealth Community Mobility Assessment, An Automated System for Early Diagnosis and Disease Progression}
%\orcidID{1111-2222-3333-4444} 

\author{Albara Ah Ramli\inst{1}\and
Xin Liu\inst{1}\and 
Erik K. Henricson\inst{2}}

\renewcommand{\thefootnote}{\alph{footnote}}
%\footnotetext[0]{* These two authors contributed equally to this work.}
\authorrunning{}%{Rex Liu et al.}
\titlerunning{Walk4Me: Telehealth Community Mobility Assessment}

\institute{
Department of Computer Science, School of Engineering\\
\and Department of Physical Medicine and Rehabilitation, School of Medicine\\
University of California, Davis\\
\email{\{arramli,xinliu,ehenricson\}@ucdavis.edu} \\
}

\maketitle              % typeset the header of the contribution

\begin{abstract}
We introduce Walk4Me, a telehealth community mobility assessment system designed to facilitate early diagnosis, severity, and progression identification. Our system achieves this by 1) enabling early diagnosis, 2) identifying early indicators of clinical severity, and 3) quantifying and tracking the progression of the disease across the ambulatory phase of the disease. To accomplish this, we employ an Artificial Intelligence (AI)-based detection of gait characteristics in patients and typically developing peers. Our system remotely and in real-time collects data from device sensors (e.g., acceleration from a mobile device, etc.) using our novel Walk4Me API. Our web application extracts temporal/spatial gait characteristics and raw data signal characteristics and then employs traditional machine learning and deep learning techniques to identify patterns that can 1) identify patients with gait disturbances associated with disease, 2) describe the degree of mobility limitation, and 3) identify characteristics that change over time with disease progression. We have identified several machine learning techniques that differentiate between patients and typically-developing subjects with 100\% accuracy across the age range studied, and we have also identified corresponding temporal/spatial gait characteristics associated with each group. Our work demonstrates the potential of utilizing the latest advances in mobile device and machine learning technology to measure clinical outcomes regardless of the point of care, inform early clinical diagnosis and treatment decision-making, and monitor disease progression.

\end{abstract}
\section{Introduction} 
The increasingly rapid growth of mobile and carry-on devices and the consequent enormous community data production opened the door to the opportunity of exploring out-of-the-lab data and obtaining a more realistic understanding of how the treatments and rehabilitations affect the patient's daily life out-of-the controlled environmental conditions. This promoted the development of advanced tools to facilitate manipulating and constructing the produced data. 

However, this opportunity has not been exploited effectively, even though several existing commercial systems target specific aspects such as disease detection, athlete assessments, etc. These systems are costly and require to be installed in hospitals or labs. There is a lack of a system that is publicly available, free, easy to use, and scalable to upgrade by researchers, developers, and clinicians. However, building such a system (that can handle big data such as steam of data, e.g., time-series data collected from sensors) which contains visualizers, recommenders, and identifiers for clinical purposes, is not a trivial task: it involves a large number of subtle decisions/features that require experience and close collaboration from the expertise, e.g., clinicians, doctors, developer, and data scientists. 

In this paper, we present Walk4me, a telehealth community mobility assessment, an automated system for early diagnosis, severity, and progression identification. This system is a tool that presents visualization to facilitate the exploration process. Walk4Me system enables both open-ended and focused exploration. It provides classical machine learning (CML) and deep learning (DL) algorithms~\cite{review,bwcnn} to help researchers engage in disease identification based on time-series data. It aims to allow researchers to answer more specific questions using ML compared to a traditional specification tool. The system automatically extracts CFs directly for the raw signals, such as the number of steps, distance, and step length. It also provides several visualizations, such as the 3d shape of the signal and the correlation between different clinical features. It also provides a feature for lining up and synchronizing the raw signal with video.

We conducted two studies using our tool for DMD patients~\cite{dmdpropose20davis,dmd} and post-stroke survivors to analyze some aspects of the disease characteristics and related data exploration, extract clinical features, utilize AI for early diagnosis purposes, and identify the most related biomechanical signals. We collected participants' data (from patients and healthy subjects) to test and develop our system and enhance our approaches. As a result, we propose a method to help facilitate the task of clinicians, researchers, and developers and help early diagnose, measure the treatment effectiveness, and better estimate the disease progression to improve the life of patients in this domain.

\section{System Design}

\subsection{System Architecture}
\begin{figure*}[h!]%[t!]
\centerline{\includegraphics[width=1\textwidth]{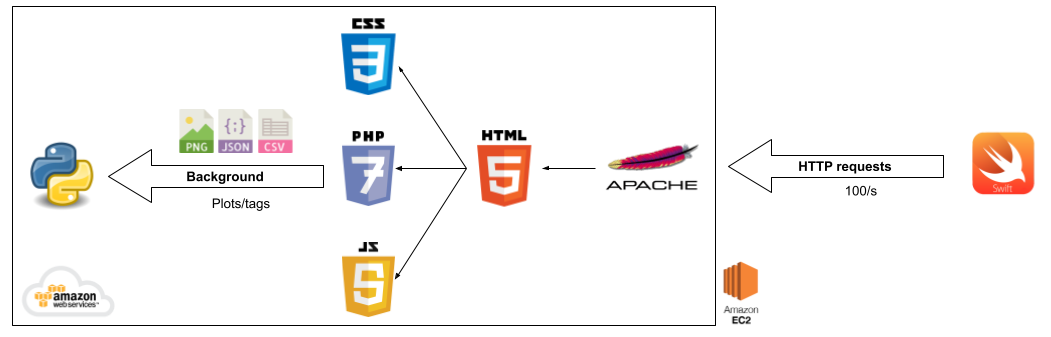}}
\caption{Project-specific login form for the Walk4Me system, enabling separate accounts for each project}
\label{walk4me:HLD}
\end{figure*}

The Walk4Me system is composed of two main components: the server-side and the client-side, as depicted in Fig.~\ref{walk4me:HLD}. The server-side technologies are used for data management, processing, and analysis, while the client-side technologies are used for data collection and transmission.

The system utilizes the following technologies:

\begin{itemize}
\item Server-side
\begin{itemize}
\item Webserver: Apache
\item Programming languages: PHP, JavaScript, and Python
\item Data storage formats: CSV and JSON
\item Computational resources: Amazon Web Services (AWS) EC2 instances
\end{itemize}
\item Client-side
\begin{itemize}
\item Devices: iPhone with an accelerometer, gyroscope, pedometer, and GPS sensors
\item Programming language: Swift
\item Data storage formats: CSV and JSON
\end{itemize}
\end{itemize}

The server-side technologies enable data processing and analysis, such as feature extraction, machine learning model training, and statistical analysis. The client-side technologies enable real-time data collection and transmission, allowing the system to collect and store data from multiple participants.

%----------------------%-
%----------------------%-
%----------------------%-
%----------------------%-

\subsection{Client side}
We developed a smartphone app to be used by the participants to collect and stream sensory data from a smartphone that is attached to the participant's body (i.e., in the DMD study, the smartphone was placed on the central mass of the body). The smartphone app has 3 phases: The off, the Ready, and the Streaming phases. In the Off phase, the app is in an off mode where no data comes in or out. In the Ready phase, the app is on standby, waiting for a signal from the server to start streaming. During the Streaming phase, the app actively streams the sensory data online to the server.
%=====================
\begin{figure*}[!t]
\centering
\centerline{

\subfigure[Off phase]{%
\label{fig:PHONE_3}
\includegraphics[height=2.5in]{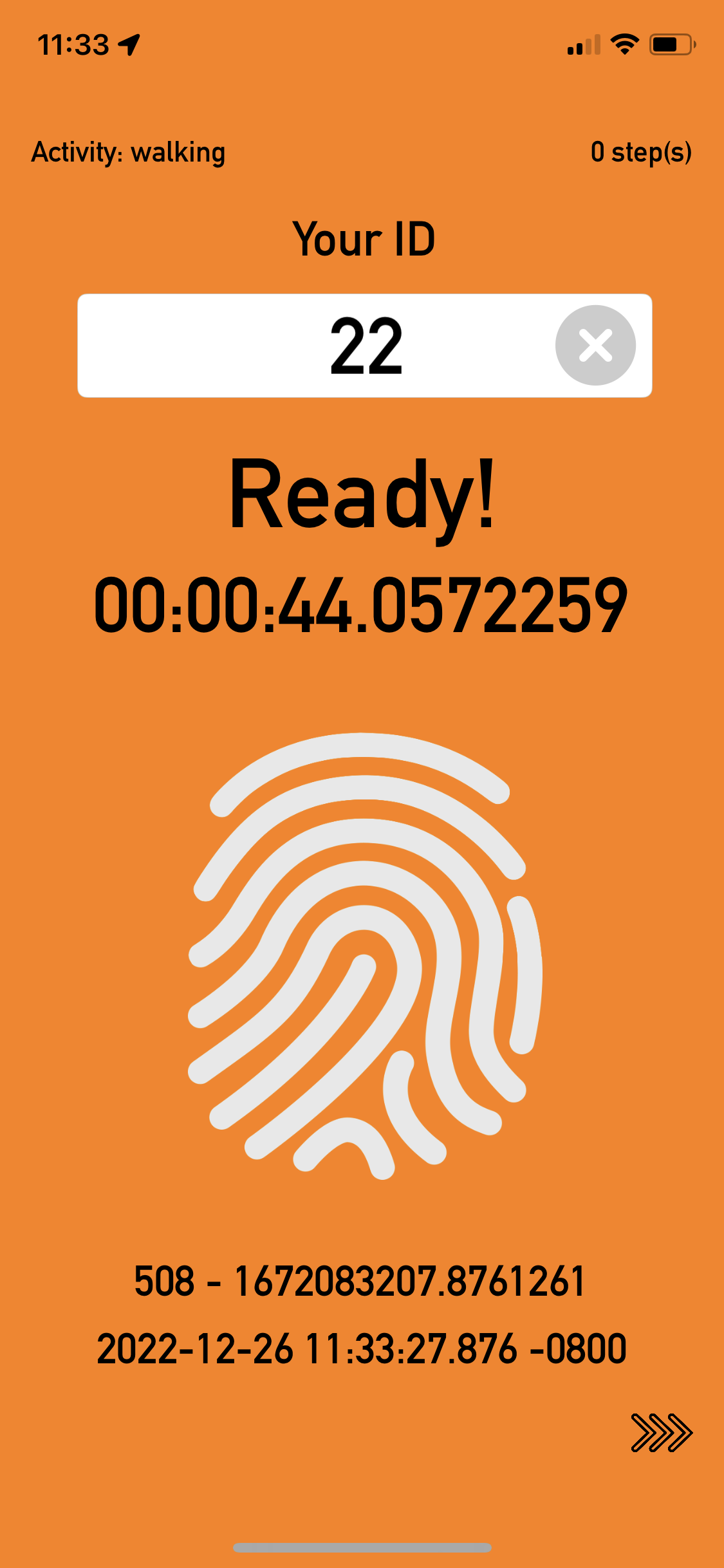}
}

\subfigure[Ready phase]{%
\label{fig:PHONE_1}
\includegraphics[height=2.5in]{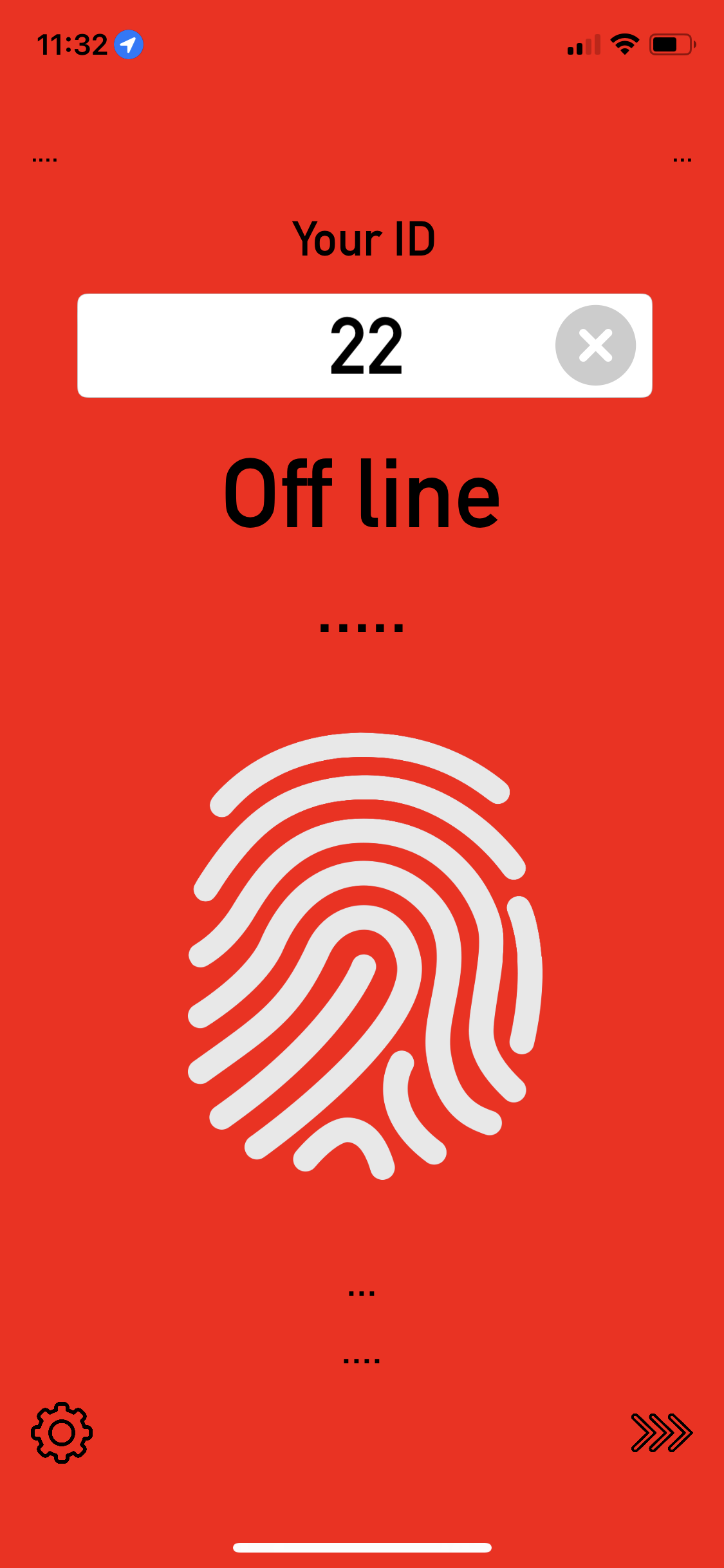}
}

\subfigure[Streaming phase]{%
\label{fig:PHONE_2}
\includegraphics[height=2.5in]{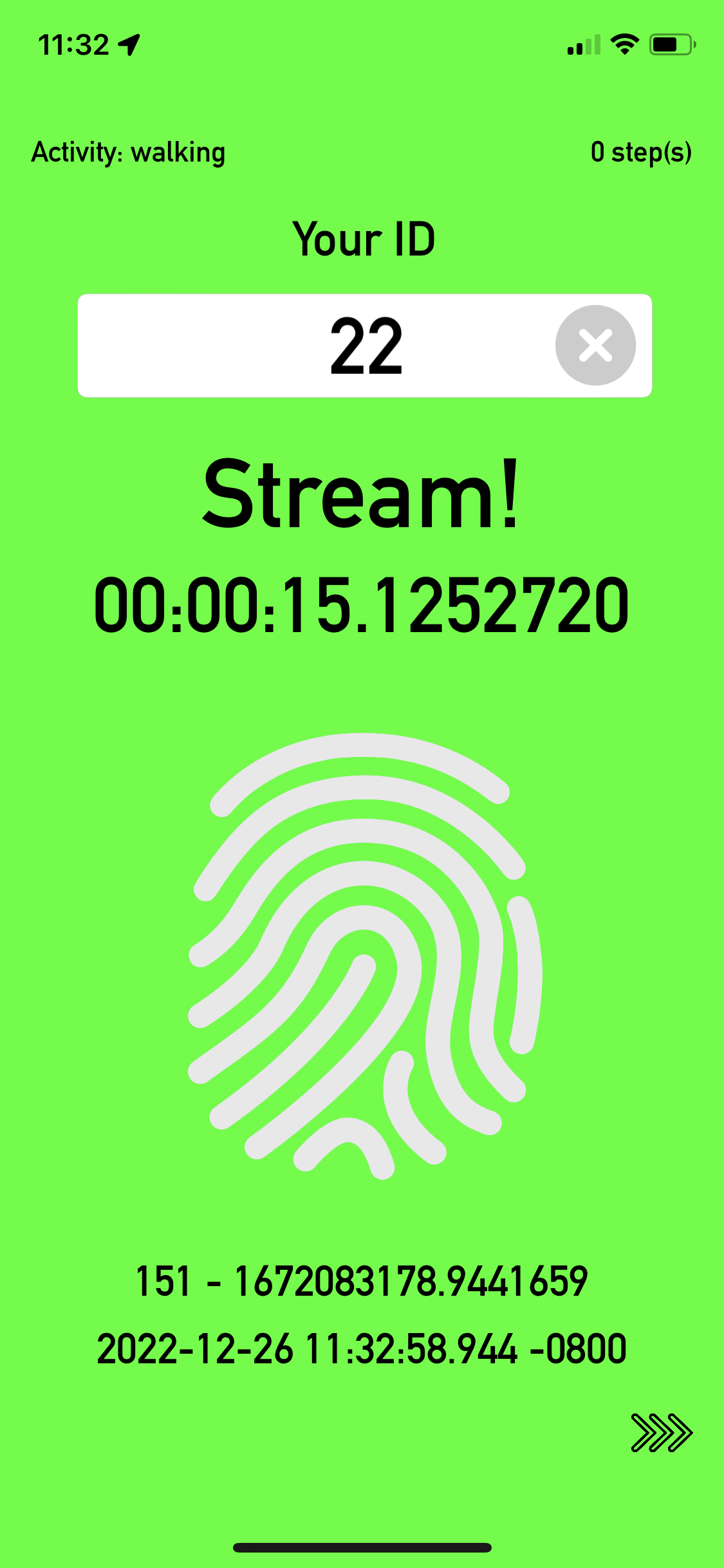}
}

}
\caption{This figure shows the Graphical User Interface (GUI) for the Walk4Me smartphone app, which has three main phases: (a) the "Off" phase, where the app is ready to stream but is not yet active; (b) "Ready" phase, where the app is inactive and waiting for the user to initiate streaming; (c) "Streaming" phase, where the app sends sensor data via the internet to the server. The GUI provides a user-friendly interface for streaming data and allows the user to monitor the status of the data transmission. The figure can help users understand the functionality of the Walk4Me app and its features.}
\end{figure*}

%============================%============================%============
%============================%============================%============

\subsection{Server side}
\begin{figure}[h!]%[t!]
\centerline{\includegraphics[width=1\textwidth]{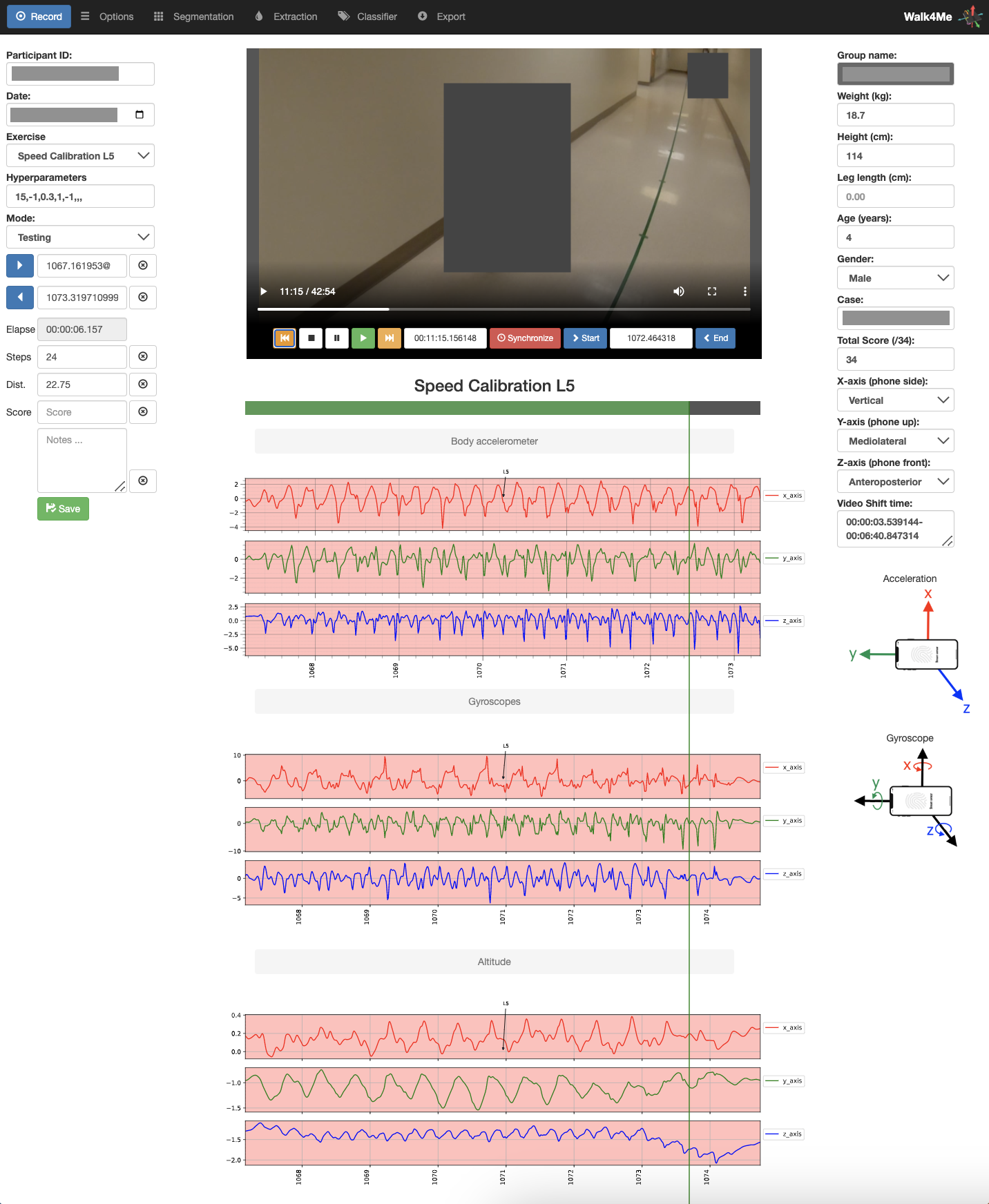}}
\caption{The web portal of the Walk4Me system. The window displays a patient's profile and the raw signal of one of the gait activities}
\label{fig:GUI_4}
\end{figure}

We developed a web portal application to facilitate access, exploration, data entry, and data upload for researchers working with the Walk4Me system. The application's server offers several essential functions, including authentication to ensure only authorized users can access the portal, initiating data collection through activating the streaming mode using a record button, browsing and visualizing collected data, synchronizing video and sensor data, marking activities performed by participants using time and demographic information, analyzing and processing data, and running different machine learning models for various tasks such as classification and regression using AI-based algorithms. Fig.\ref{fig:GUI_4} shows the web portal interface.
%----------------------%-
%----------------------%-
%----------------------%-
%----------------------%-

\subsubsection{Authentication login}
\begin{figure}[h!]%[t!]
\centerline{\includegraphics[width=0.7\textwidth]{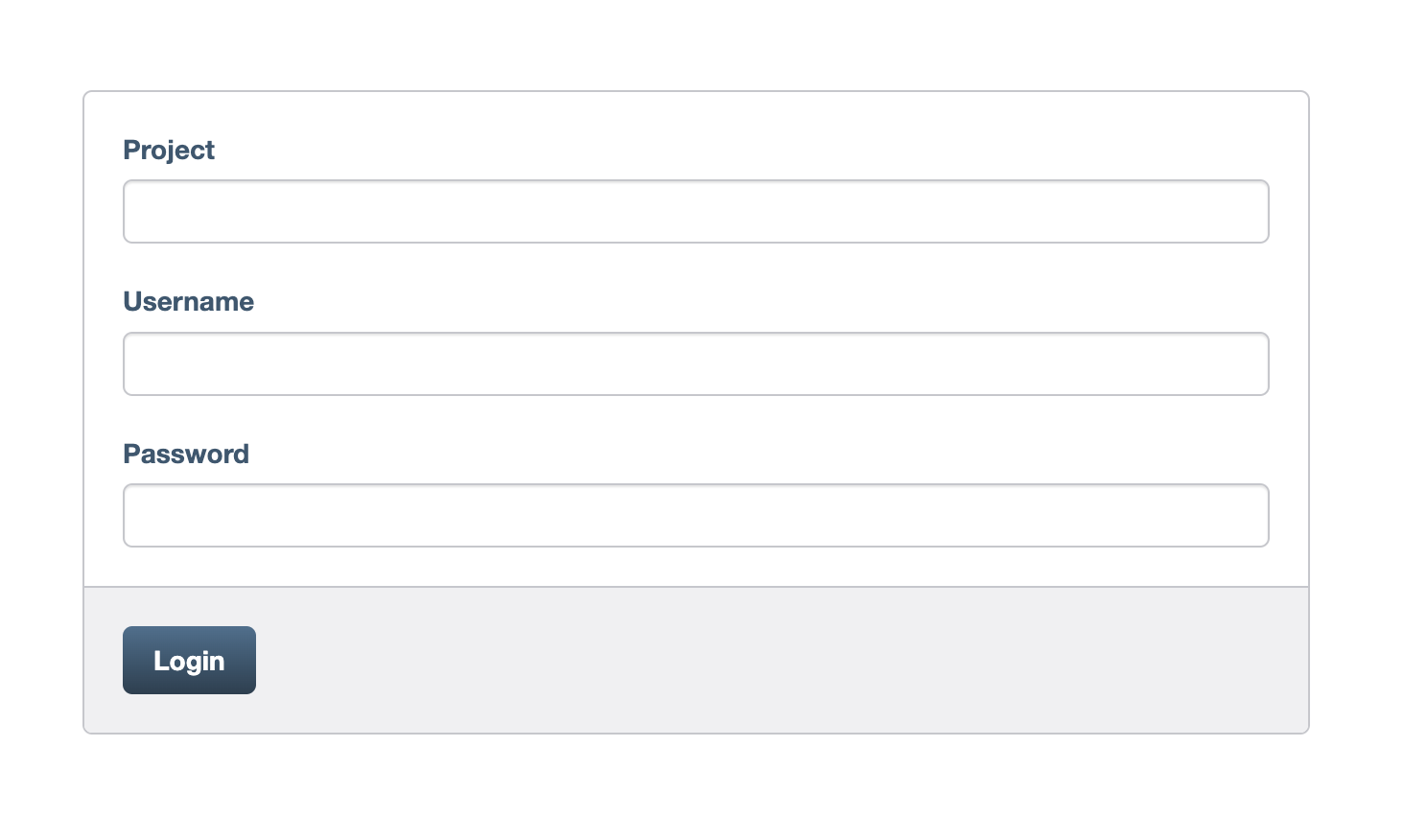}}
\caption{This figure shows the login form of the Walk4Me system, which allows for the creation of separate accounts for each project.}
\label{fig:LOGIN}
\end{figure}

Fig.~\ref{fig:LOGIN} shows the login page of the web application, which allows researchers to access their project data securely. The system provides separate logins for each project, enabling multiple groups of researchers to work in their own private space without accessing data from other projects.
%----------------------%-
%----------------------%-
%----------------------%-
%----------------------%-
\section{Feature Analysis and Implementation}

The Walk4Me system provides a range of tools for exploring and analyzing sensory data. These tools include feature extraction, data visualization, and statistical analysis. By using these tools, researchers can gain valuable insights into the gait patterns of patients and develop more effective treatments for mobility impairments. Additionally, the system offers the ability to implement these features in machine learning models for automated classification and prediction.
\subsection{Feature Extraction}

The system provides a feature extraction tool that can extract a variety of gait features from the accelerometer raw signal. These features include both clinical and computational features, such as the number of steps, total distance, step lengths, average step length, step duration, total duration, step frequency, and average speed. These features are useful for data analysis and can be used to train classical machine learning classifiers.

\subsection{Orientation}

%----------------------%-
%----------------------%-
\begin{figure}[h!]%[t!]
\centerline{\includegraphics[width=0.4\textwidth]{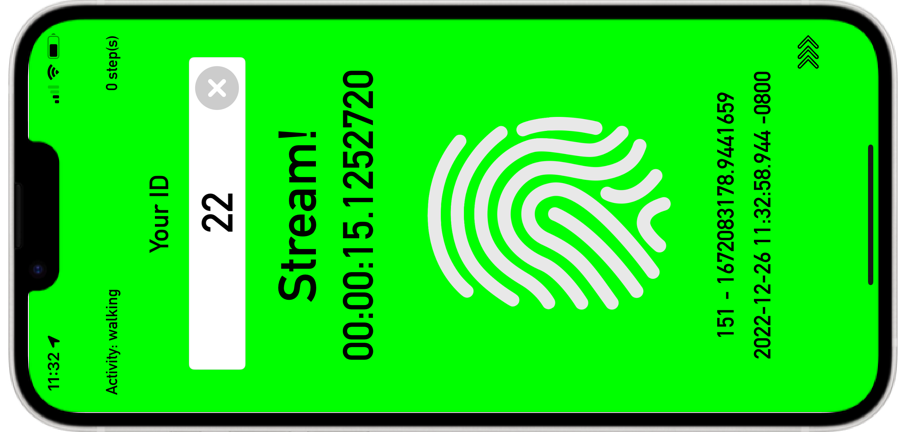}}
\caption{Phone orientation during streaming data.}
\label{fig:new_iphone}
\end{figure}
%----------------------%-
%----------------------%-

\subsection{Dashboard}

The clinical dashboard is another tool provided by our system that enables you to illustrate the measurements obtained using signal processing. For instance, you can create a visualization for the number of steps at different step velocities, a distribution histogram of the number of steps at different step velocities, step length per forward acceleration peaks, and the distance traveled by step length.

%----------------------%-
%----------------------%-
\begin{figure}[ht!]%[t!]
\centerline{\includegraphics[width=1\textwidth]{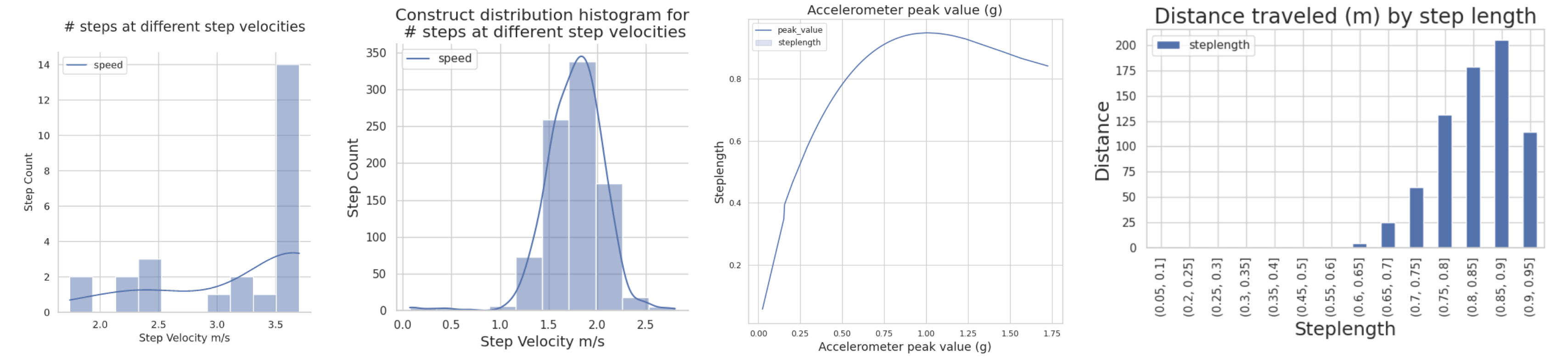}}
\caption{This figure shows visualizations provided by the system, including the number of steps at different step velocities, distribution histogram of the number of steps at different step velocities, step length per forward acceleration peaks, and the distance traveled by step length. These visualizations are part of the data analysis process that the system provides to users. }
\label{fig:dashboard2}
\end{figure}
%----------------------%-
%----------------------%-

%----------------------%-
%----------------------%-
\begin{figure}[ht!]%[t!]
\centerline{\includegraphics[width=1\textwidth]{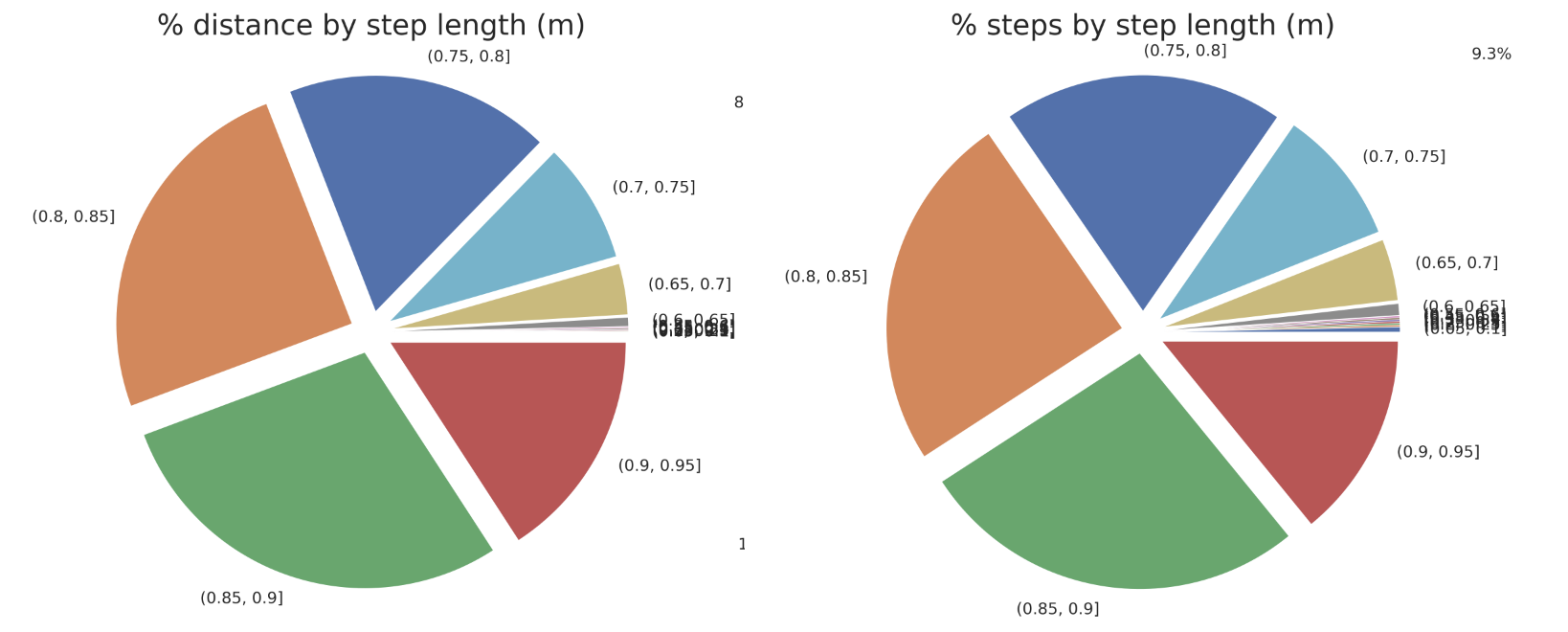}}
\caption{This figure shows the percentage of distance and the number of steps per step length.}
\label{fig:dashboard1}
\end{figure}
%----------------------%-
%----------------------%-

%----------------------%-
%----------------------%-
\begin{figure}[ht!]%[t!]
\centerline{\includegraphics[width=1\textwidth]{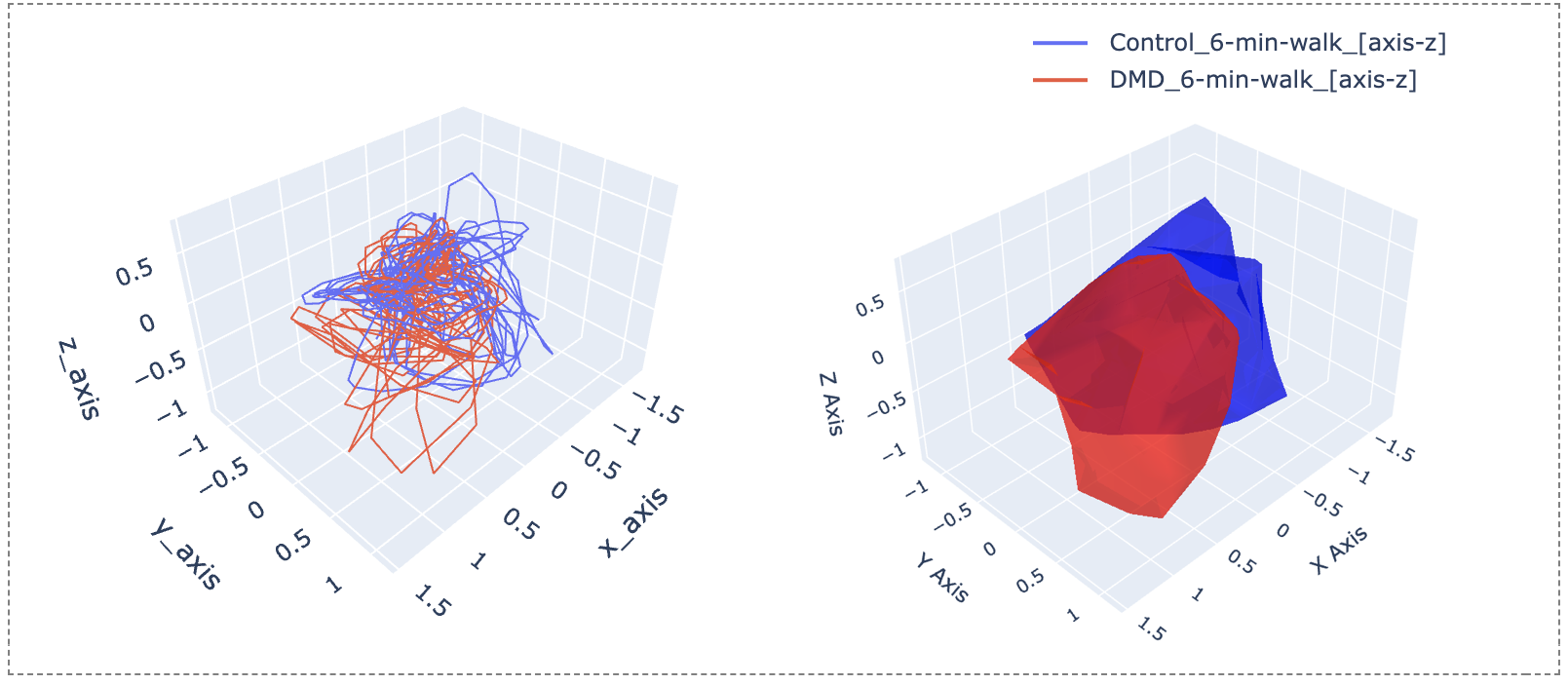}}
\caption{3D visualization of forward acceleration for a participant with DMD compared to a TD peer.}
\label{fig:dashboard3}
\end{figure}
%----------------------%-
%----------------------%-

\subsection{Signal overlap}
\begin{figure}[ht!]%[t!]
\centerline{\includegraphics[width=1\textwidth]{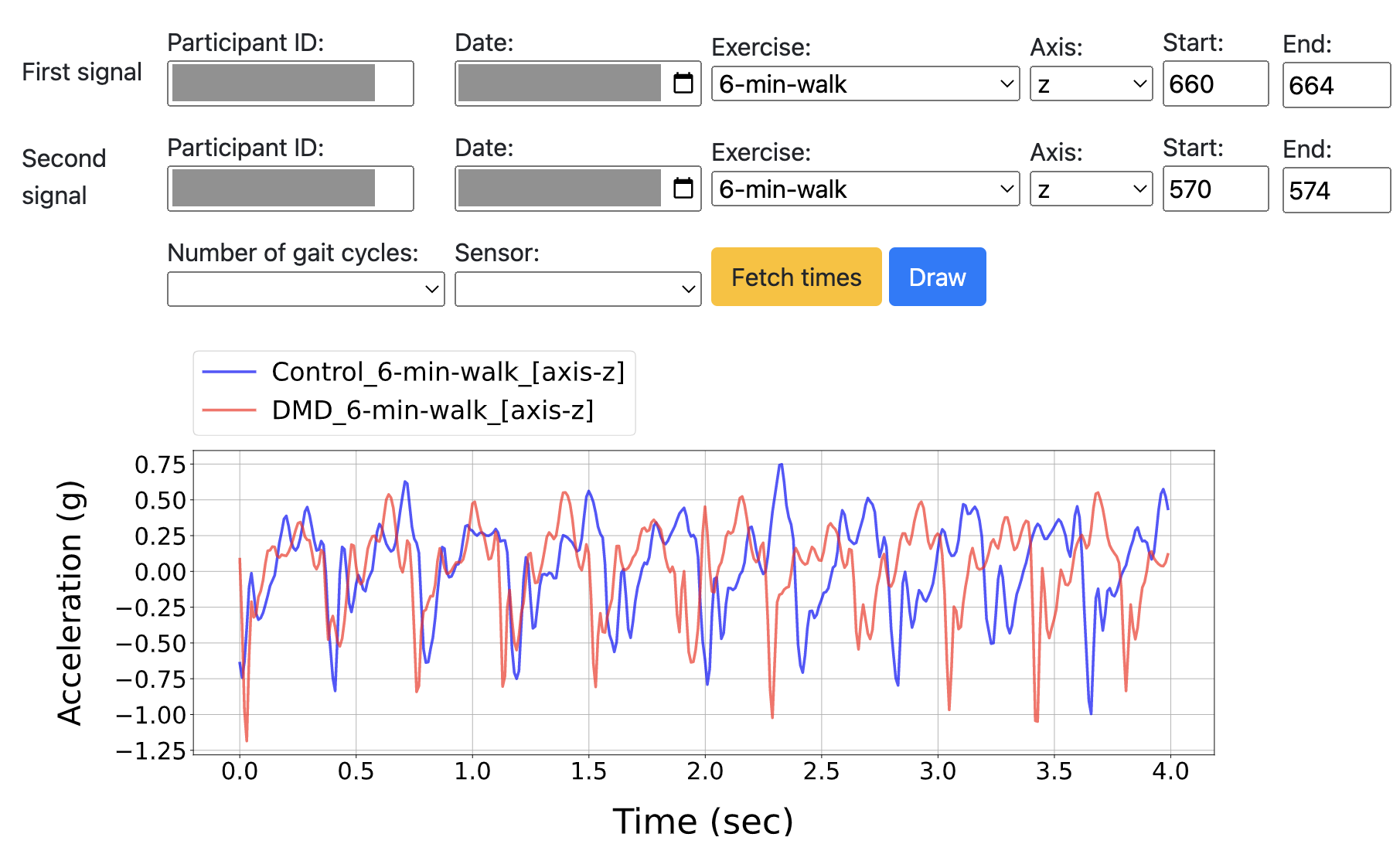}}
\caption{Comparison of two raw accelerometer signals recorded during the 6-minute walk test (6MWT) performed by a participant with Duchenne muscular dystrophy (DMD) and typically developing peer (TD). The graph shows the acceleration data from the forward movement overlapped to visualize the differences in gait patterns between the two groups.}
\label{fig:overlap}
\end{figure}

A signal overlap is a tool provided to the researchers to compare two signals either in the same patient or across different participants. One of the benefits of this tool is comparing a gait cycle of a control participant vs. a patient with a mobility disorder. Fig.~\ref{fig:overlap} shows the tool's user interface. Nothing to mention that this tool also allowed researchers to mark a specific time event. For example, in walking analysis, researchers can mark different gait events such as Initial contact, Opposite Toe off, Heel rise, Opposite initial contact, Toe off, Feet adjacent, Tibia vertical, and Next initial contact.

%----------------------%-
%----------------------%-
%----------------------%-
%----------------------%-

\section{Data Processing}
The system provides three stages of data processing: feature extraction, raw data, and classification with dimensionality reduction.

\section{Machine Learning Tool}
The system includes a patient profile for each participant, which contains demographic and characteristic data. Additionally, researchers can label patients to indicate their class for use in the ML classifier. There is no limit on the number of classes that can be used. In binary classification, patients are labeled with two classes. For example, in the DMD study, patients were labeled as either DMD or control. In the stroke study, patients were labeled as either Stroke vs. control, or More vs. Less affected side.

Our system provides a machine learning (ML) tool to train the classification model. The ML tool includes two types of ML training methods: classical machine learning using the scikit-learn library, and deep learning using TensorFlow. With this tool, users can choose the best machine learning model and method for their specific dataset and classification task, enabling high accuracy and robustness in their machine learning models.

\subsection{Classical Machine Learning}
In classical machine learning, our system utilizes the generated classification features to train a variety of different classifiers. Specifically, we provide support for six different classifiers: AdaBoost, Random Forest, Bagging (ensemble), Gradient Boosting, Decision Tree, Support Vector Machine (SVM), K-Nearest Neighbors, Gaussian Naive Bayes, and Logistic Regression. This enables users to choose the best classifier for their particular dataset and classification task, and achieve high accuracy and robustness in their machine learning models.

\subsubsection{Feature correlation}
\begin{figure}[ht!]%[t!]
\centerline{\includegraphics[width=\textwidth]{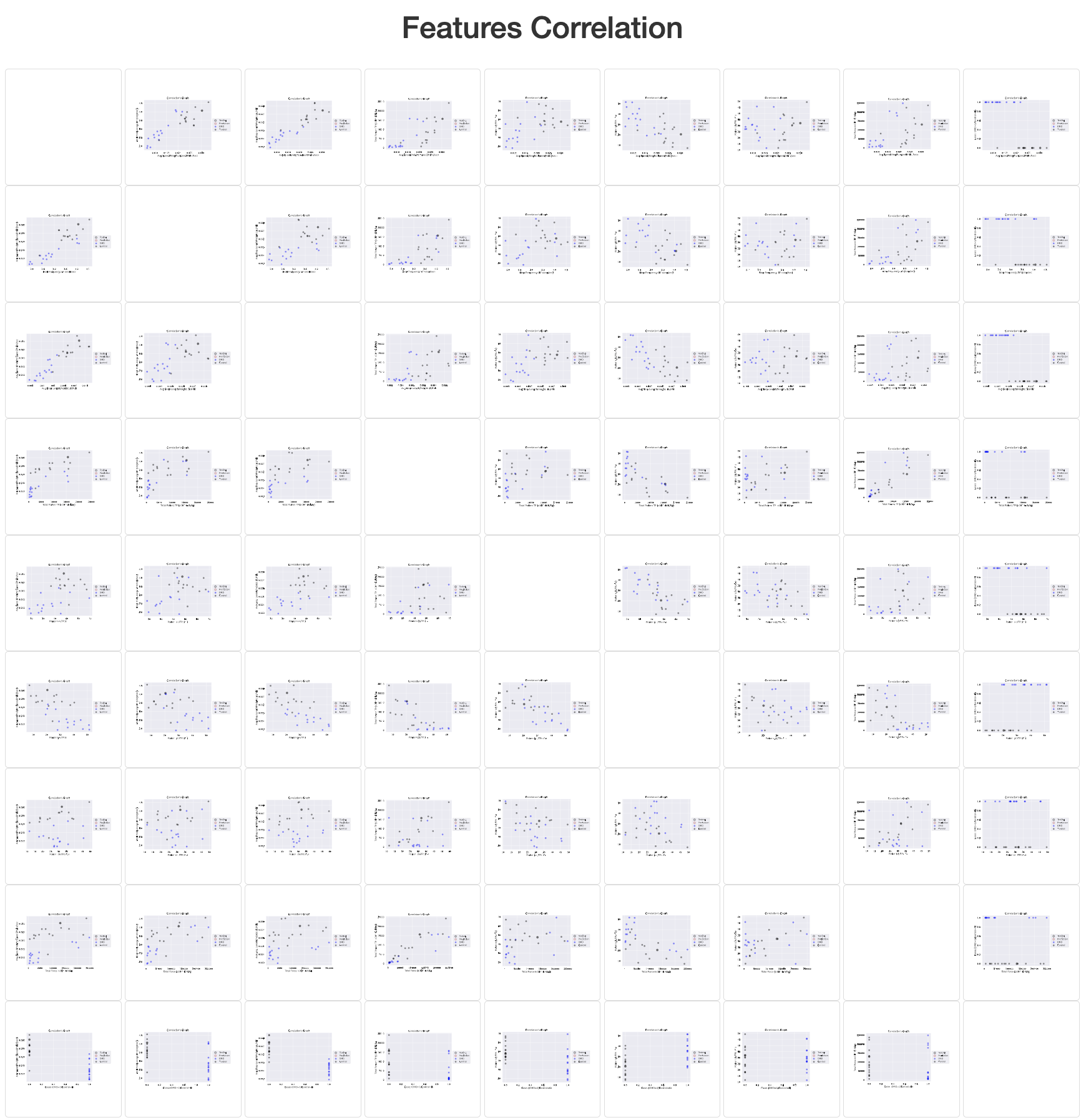}}
\caption{This figure shows the correlation between features across 8 of the clinical features (CF) in the DMD study. The figure provides insight into how these features may be related and can help guide data analysis and feature selection for machine learning models.}
\label{fig:CML-CF_2}
\end{figure}

The system provides a visual representation of the CF as shown in Fig.\ref{fig:CML-CF_2}
%----------------------%-
%----------------------%-
%----------------------%-
%----------------------%-

\subsubsection{Dimensionality Reduction}
\begin{figure}[ht!]%[t!]
\centerline{\includegraphics[width=\textwidth]{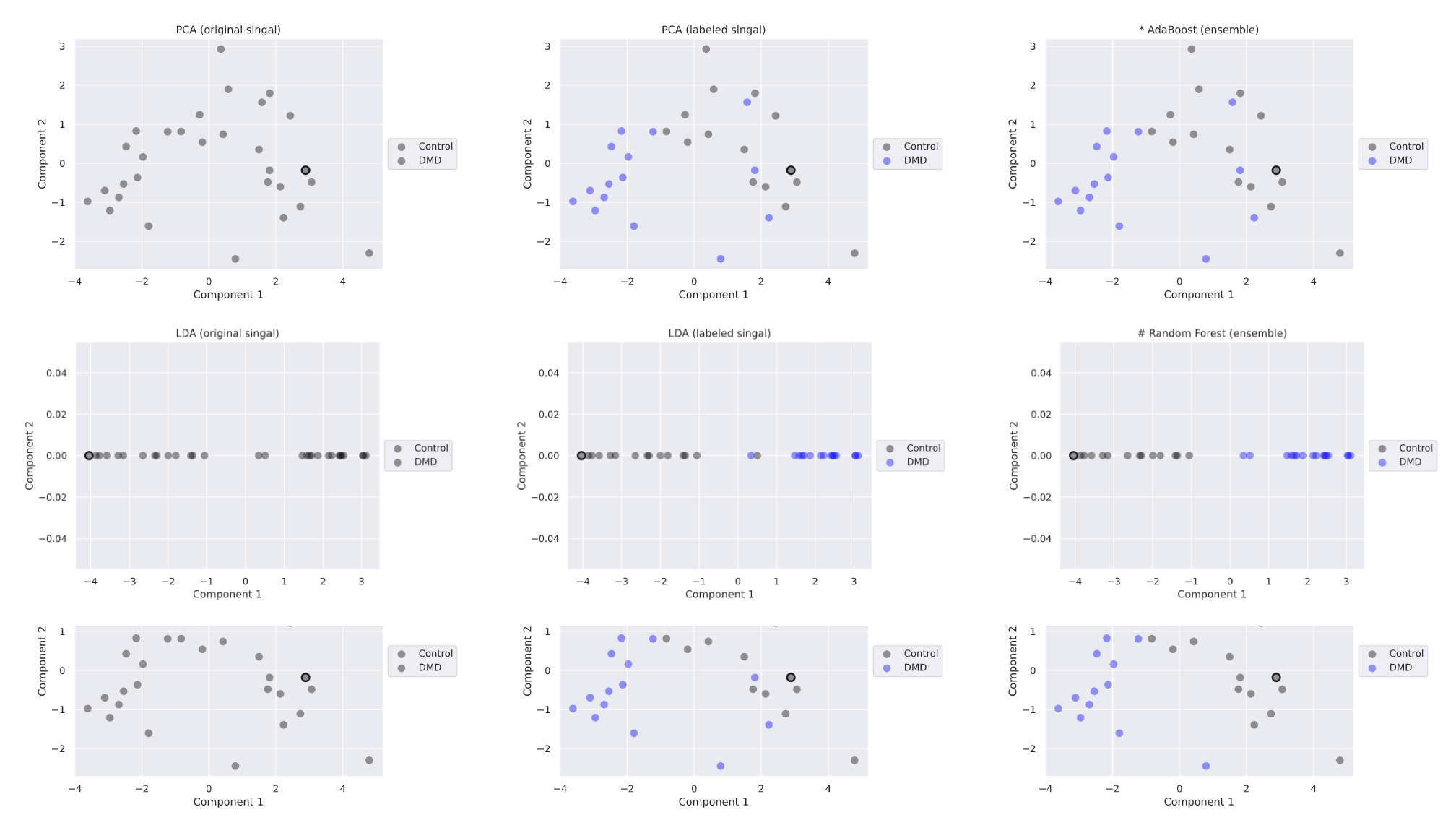}}
\caption{This figure illustrates the PCA and LDA representation of patient data. It includes the original data points, ground truth labels, and the corresponding classification results obtained by the system (multiple classifiers are used, and this is an example). The visualization provides valuable insights into the classification performance of the system.}
\label{fig:CML-CF_0}
\end{figure}

Our system implements dimensionality reduction as an optional step before the classifier to reduce the feature space and improve model performance. Two methods of dimensionality reduction are provided: Linear Discriminant Analysis (LDA) and Principal Component Analysis (PCA). Figure \ref{fig:CML-CF_0} shows an example of implementing LDA and PCA with an AdaBoost binary classifier on a participant. This step can greatly improve the efficiency and accuracy of the classification process, particularly when dealing with high-dimensional datasets.
%----------------------%-
%----------------------%-
%----------------------%-
%----------------------%-

\subsection{Deep Learning for Participant Classification}
Our system offers a powerful tool for training deep learning models to classify participants. The interface allows for flexible tuning of hyperparameters, including the number of epochs and the use of leave-one-out and ensemble learning techniques. This enables users to optimize their models and achieve high accuracy in the participant classification task.

\section{Conclusion}
This work presents Walk4me, a telehealth system that uses Artificial Intelligence (AI) to detect gait characteristics in patients and typically developing peers. The system collects data remotely and in real-time from device sensors, extracts gait characteristics and raw data signal characteristics, and uses machine learning techniques to identify patterns that can facilitate early diagnosis, identify early indicators of clinical severity, and track disease progression across the ambulatory phase of the disease. The authors have identified several machine learning techniques that differentiate between patients and typically-developing subjects with 100\% accuracy across the age range studied and have identified corresponding temporal/spatial gait characteristics associated with each group. The system has the potential to inform early clinical diagnosis, treatment decision-making, and monitor disease progression.


\bibliographystyle{unsrt}
\bibliography{reference}

\begin{thebibliography}{1}

\bibitem{review}
Rex Liu, Albara~Ah Ramli, Huanle Zhang, Erik Henricson, and Xin Liu.
\newblock {An Overview of Human Activity Recognition Using Wearable Sensors:
  Healthcare and Artificial Intelligence}.
\newblock In Bedir Tekinerdogan, Yingwei Wang, and Liang-Jie Zhang, editors,
  {\em Internet of Things -- ICIOT 2021}, pages 1--14, Cham, 2022. Springer
  International Publishing.

\bibitem{bwcnn}
Albara~Ah Ramli, Rex Liu, Rahul Krishnamoorthy, I.~B. Vishal, Xiaoxiao Wang,
  Ilias Tagkopoulos, and Xin Liu.
\newblock Bwcnn: Blink to word, a real-time convolutional neural network
  approach.
\newblock In {\em Internet of Things - ICIOT 2020}, pages 133--140, Cham, 2020.
  Springer International Publishing.

\bibitem{dmdpropose20davis}
Albara~Ah Ramli, Alina Nicorici, Poonam Prasad, JaiHui Hou, Craig McDonald, Xin
  Liu, and Erik Henricson.
\newblock An automated system for early diagnosis, severity, and progression
  identification in duchenne muscular dystrophy: a machine learning and deep
  learning approach.
\newblock In {\em Annual Human Genomics Symposium -- University of California
  Davis Medical Center}, pages 12--12, 2020.

\bibitem{dmd}
Albara~Ah Ramli, Huanle Zhang, Jiahui Hou, Rex Liu, Xin Liu, Alina Nicorici,
  Daniel Aranki, Corey Owens, Poonam Prasad, Craig McDonald, and Erik
  Henricson.
\newblock {Gait Characterization in Duchenne Muscular Dystrophy (DMD) Using a
  Single-Sensor Accelerometer: Classical Machine Learning and Deep Learning
  Approaches}.
\newblock arXiv, 2021.

\end{thebibliography}

\end{document}